\documentclass[12pt]{article}
\usepackage{amsfonts,euscript,srcltx}
\usepackage{cite}
\begin{document}

\title{\bf  Massless Chiral Supermultiplets of Higher Spins and the $\theta$-Twistor}
\author 
{ ~ M. Chaichian, ${}^{ {a}}$ ~ A. Tureanu ${}^{{a}}$ and 
A.A.~Zheltukhin ${}^{b,c,d}$
\\
{\normalsize ${}^a$  Department of Physics, University of Helsinki, 
}\\
{\normalsize and
 Helsinki Institute of Physics, P.O. Box 64, FIN-00014 Helsinki, Finland,
}\\
{\normalsize ${}^b$ 
Kharkov Institute of Physics and Technology, 61108 Kharkov, Ukraine,}\\
{\normalsize ${}^{c}$ 
 Department of Physics, Stockholm University, 10 691 Stockholm, Sweden,}\\
{\normalsize ${}^{d}$  
 NORDITA, Roslagstullsbacken 23, 106 91, Stockholm, Sweden} 
}                                            
\date{}
\maketitle
\begin{abstract}

Recently N. Berkovits, motivated by the supertwistor description of 
${\cal N}=4 \,\, D=4$ super Yang-Mills, considered the 
generalization
 of the
 ${\cal N}=1 \,\, D=4 \, \, \, \theta$-twistor construction 
to $ D=10$ and applied it for a compact covariant description of 
${\cal N}=1 \,\, D=10$ super Yang-Mills. This supports the relevance 
of the $\theta$-twistor as a supersymmetric twistor alternative to 
the well-known supertwistor. The minimal breaking of superconformal 
symmetry is an inherent  property of the $\theta$-twistor received 
from its fermionic components, described by 
 a Grassmannian vector instead of a Grassmannian scalar in the supertwistor. 
The $\theta$-twistor description of the  ${\cal N}=1 \,\, D=4$ massless chiral 
supermultiplets $(S, S + 1/2)$ with spins $S=0,1/2,1,3/2,2,...$ 
 is considered here. 
The description permits to restore the auxiliary $F$ fields of the chiral 
supermultiplets absent in the supertwistor approach.
 The proposed formalism is naturally generalized to
${\cal N}=4 \,\, D=4 $ and can be used for an off-shell description of 
the corresponding super Yang-Mills theory.

\end{abstract}

\section{Introduction}

The supertwistor is a supersymmetric generalization of the Penrose
 twistor \cite{PR} constructed by Ferber at the end of the 1970s \cite{Fbr,Witt1}. The recent discovery of the supertwistor role in computing 
 ${\cal N}=4 \,\, D=4$ multigluon amplitudes in super Yang-Mills theory 
\cite{Witt2,Nai,Ber,RSV,GKh} and their connection with the twistor 
strings strongly extended the range of the previous applications of 
the Penrose twistor program (see, e.g.,
\cite{ADHM,CFGT,Shir,BC,Sieg1,Sieg2,BZ,UZ}). 
 Concerning the role of the supertwistor in superstring and super p-brane theory we recall that their use made possible the covariant 
quantization of tensionless superstrings and  supermembranes, as 
well as the proof of the absence of the critical
 dimension in their quantum description.   
After this progress, the supersymmetric twistor variables had been 
used to construct the twistor-like Lagrangian and Hamiltonian for 
the $ D=10$ Green-Schwarz superstring (and  $ D=11$  supermembrane), 
 to solve the problem of the Lorentz covariant splitting of its first and 
second class constraints and to find the irreducible covariant 
realization of the $\kappa$-symmetry generators  \cite{BZ_GS}. 
However, the covariant classical BRST charge of the superstring, 
derived in the  approach, has turned out to be such a complicated 
function of the canonical variables that blocked the transition to a 
quantum BRST operator. The twistor transform of the superstring 
action \cite{BZ_GS} considered in \cite{Uvar_GS} presented it in the 
supertwistor form with the $D=10$ supertwistor realized as the 
fundamental representation of the $OSp(32|1)$ supergroup  
\cite{VHVP}. This supergroup realizes the superconformal 
transformations like the superconformal group $SU(2,2|1)$ in the 
$D=4$ Minkowski space extended by the Grassmannian spinor 
coordinates $\theta$.
 Some promising tools towards the solution of this problem are connected with the
 approach \cite{Ber_cq} using the  $D=10$ pure spinors, previously considered in 
\cite{Nils,Howe,Ton}, in the role of the discussed
 twistor-like variables (see also \cite{BCher}).
The approach made possible computing the superstring amplitudes 
\cite{Ber_ampl}, however its relation with the Green-Schwarz 
superstring is still open. These and other known results show the 
important role of the concept of twistor in the superstring and 
Yang-Mills theories, and stimulate its further
 development accompanied by unification with supersymmetry and higher dimensions. 

In the recent paper \cite{Ber_10}, Berkovits proposed to generalize
the well-known supertwistor description of the  ${\cal N}=4 \,\, 
D=4$ super Yang-Mills theory to the case ${\cal N}=1 \,\, D=10$ 
super Yang-Mills theory
 to get its compact covariant description.
The proposal was partially stimulated by the papers 
\cite{Wit_10,STVZ,Ber_10a}, showing many similar features between 
Yang-Mills
 and superstring theories in $D=10$.
 In \cite{Ber_10} were discussed new $D=10$ 
super-twistor variables $Z=(\lambda^{\alpha}, \mu_{\alpha}, \Gamma^{m})$,
 where $\lambda^{\alpha}$ and  $\mu_{\alpha}$ are constrained spinors
 and $\Gamma^{m}$ is the Grassmannian 10-vector 
$\Gamma^{m}=(\lambda\gamma^{m}\theta)$ substituted for the Grassmannian scalar 
$\eta=(\lambda\theta)$ used in the supertwistor.
 It results in a relation between the scalar superfield $\Phi(Z)$ and the 
$ D=10$ superfield Yang-Mills vertex operator 
$\lambda^{\alpha}A_{\alpha}(x,\theta)$  of the pure spinor 
superstring formalism.  In addition, the cubic super Yang-Mills 
amplitude was found to be proportional to the integral of 
$\Phi^3(Z)$ over the $Z$-space. The observations  \cite{Ber_10} shed 
a new light on the connection between the GS and RNS strings with 
the  $D=10$ super Yang-Mills theories.  

The super-twistor variables $Z=(\lambda^{\alpha}, \mu_{\alpha}, \Gamma^{m})$ 
discussed in \cite{Ber_10} are the generalization to the case $D=10$ of 
the $D=4 \,\, {\cal N}=1$  supersymmetric $\theta$-twistor $\bf\Xi$ studied 
in \cite{Z}. The  $\bf\Xi$  components are presented by the triple 
 ${\bf\Xi_{\cal A}}=(-il_{\alpha},\, \bar\nu^{\dot\alpha},\, 2\sqrt{2}\, \bar\eta_{m})$ 
 including the known Penrose's chiral spinor $\bar\nu^{\dot\alpha}$,
accompanied by the new $D=4$ Weyl spinor $l_{\alpha}$ and the complex 
Grassmannian 4-vector $\bar\eta_{m}$. 
The $\theta$-twistor components $l_{\alpha}$  and $\bar\eta_{m}$ 
are defined by the general solution of the real null constraint 
${\bf\Xi_{\cal A}}{\bf\bar\Xi^{\cal A}}=0$, where 
 ${\bf\bar\Xi^{\cal A}}\equiv({\bf\Xi}_{\cal A})^*=
(\nu^{\alpha},\, i{\bar l}_{\dot\alpha},\, 2\sqrt{2}\, \eta^{m})$ is 
the complex conjugate of ${\bf\Xi_{\cal A}}$.
 The solution of the constraint fixes $l_{\alpha}$  and $\bar\eta_{m}$  in the form 
$$
l_{\alpha}=y_{\alpha\dot\alpha}\bar\nu^{\dot\alpha},
 \, \, \,
\bar\eta_{m} = -\frac{1}{2}(\theta\sigma_{m}\bar\nu),\, \, \,
y_{\alpha\dot\alpha}=x_{\alpha\dot\alpha}-2i\theta_{\alpha}\bar\theta_{\dot\alpha},
$$
where $y_{\alpha\dot\alpha}\equiv y_{m}\sigma_{\alpha\dot\alpha}^{m}$
 and $\theta_{\alpha}$
are known coordinates of the chiral superspace $(y_{m},\, 
\theta_{\alpha})$  \cite{WB}. The vector  $\bar\eta_{m}$ and its 
complex conjugate $\eta_{m}=(\bar\eta_{m})^* 
=-\frac{1}{2}(\nu\sigma_{m}\bar\theta)$ form the real $D=4$ 
Grassmannian vector $ \psi_{m}= \eta_{m} + \bar\eta_{m}= 
-\frac{1}{2}(\bar\nu\gamma_{m}\theta)$ firstly introduced in 
\cite{VZ} and later used for  $D=10$  in \cite{STVZ} and denoted by 
$\Gamma_{m}$ in \cite{Ber_10}. The name $\theta$-twistor used in 
\cite{Z}  for the triple  ${\bf\Xi_{\cal A}}$ 
 was motivated by its difference from the supertwistor \cite{Fbr}.
The difference is that the  $\theta$-twistor
 is covariant only under transformations of the maximal subgroup of the 
 $D=4$ superconformal group $SU(2,2|1)$ including the super-Poincar\'e group,
dilatation together with the phase and the axial $\gamma_{5}$ 
transformations, as it was shown in \cite{Z}. The broken symmetries 
of the $\theta$-twistor superspace turn out to be the superconformal 
boosts. Taking into account that both the $D=10$ super Yang-Mills 
theory and the $D=10$ $\theta$-twistor are not  superconformally 
covariant, it seems instructive to study the structure of the ${\cal 
N}=1\, \, D=4$
 massless superfields $F(\bf\Xi)$ in the superspace created by the
 $\theta$-twistor $\bf\Xi$. This is the main aim of the present paper.

First let us make some geometric comments, explaining the  origin of 
the $\theta$-twistor
 and how it differs from the supertwistor. 
 The supertwistor is the projective triple including two commuting spinors 
and the additional Grassmannian Lorentz scalar 
$\eta=\nu^\alpha\theta_\alpha$ representing the Grassmannian 
component of the supertwistor contributed by the spinor coordinate 
$\theta_\alpha$ of the $D=4\, \,  {\cal N}=1$ superspace. Using the 
projection $\eta$ instead of  $\theta_{\alpha}$ reduces the spin 
structure represented by the supertwistor and, as a result, the 
massless chiral
 supermultiplets loose their auxiliary $F$-field, yielding the on-shell
 supersymmetry transformations discussed in \cite{Fbr}.
The restriction may be overcome by the transition to the 
$\theta$-twistor\footnote{For simplicity, we discuss the  $N=1$  
supersymmetry, but the $\theta$-twistor
 generalization for the  $SU(N)$ symmetry group is automatically achieved by
 the substitution of $\theta_{\alpha}^{i}$ for  $\theta_{\alpha}$,
 where $i$ is the fundamental representation index.}. 
Contrary to the supertwistor, the Grassmannian components of the 
$\theta$-twistor
 are represented by the composite Grassmannian (or Ramond) vector $(\theta\sigma_{m}\bar\nu)$ 
  \cite{VZ} composed of the spinors $\theta_{\alpha}, \,\,\bar\nu_{\dot\alpha}$.

There is a very simple algebraic reason for the existence of the 
projective $\theta$-twistor. For a given Weyl spinor 
$\theta_{\alpha}$ with a fixed chirality one can match it either 
with the same chirality Penrose's spinor $\nu^{\alpha}$, or with the 
opposite chirality spinor ${\bar\nu}_{\dot\alpha}$. In the first 
case we obtain the Lorentz scalar $(\theta\nu)$, and respectively 
the Lorentz
 vector $(\theta\sigma_{m}\bar\nu)$ in the second case.  
The first possibility results in the supertwistor, while the second 
leads to the 
 $\theta$-twistor. 
The supertwistor components additional to the Penrose spinor 
 $\nu_{\alpha}$ are produced by the {\it left projection} of the chiral superspace coordinates
$( y_{\alpha\dot\alpha},\,\theta_{\alpha})$ on $\nu^{\alpha}$. The 
two projections form the {\it double}  $\left(\frac{1}{2}, 0\right)$ 
whose unification with $\nu^{\alpha}$ yields the projective {\it 
triple} linearly realizing  the supersymmetry transformations
 and called supertwistor \cite{Fbr}.
Because of the {\it non-Hermicity} of the matrix  
$y_{\alpha\dot\alpha}$, formed by the chiral
 coordinates,  we observe the alternative possibility to extend the chiral superspace 
 by the {\it right multiplication} of its coordinates on
the c.c. spinor $\bar\nu^{\dot\alpha}$. This way yields a new 
projective  {\it triple}, formed  by the {\it double} 
$\left(\frac{1}{2}, 1\right)$  and $\bar\nu^{\dot\alpha}$,  called 
the $\theta$-twistor \cite{Z} and forming a new
 linear representation of the supersymmetry. 
The vector component of the new {\it double}  $\left(\frac{1}{2}, 
1\right)$ is $(\theta\sigma_{m}\bar\nu)$ and it represents the 
Grassmannian component of the $\theta$-twistor. Thus, the two  
different {\it doubles} 
 $\left(\frac{1}{2}, 0\right)$ and $\left(\frac{1}{2}, 1\right)$ suggest two independent 
supersymmetric generalizations of the bosonic Penrose twistor called 
the supertwistor and the $\theta$-twistor, respectively. The 
$\theta$-twistor and the supertwistor turn out  to be general 
solutions of different supersymmetric
 constraints  \cite{Z}, generalizing the standard chirality constraint 
to superspaces extended either by $\nu_\alpha$ or 
$\bar\nu_{\dot\alpha}$. The Grassmannian vector $\bar\eta_{m}= 
-\frac{1}{2}(\theta\sigma_{m}\bar\nu)$ plays the  role of the 
Grassmannian scalar $\eta=\nu^{\alpha}\theta_{\alpha}$ in the 
superfields $F(\bf\Xi)$ on the $\theta$-twistor space. The 
$\bar\eta_{m}$ expansion of $F(\bf\Xi)$ generates a complete set of  
component fields contrary to the supertwistor set, produced by the 
$\eta$ expansion. It is because $\eta^2=0$, but $\bar\eta_{m}^2$ is 
proportional to $\theta^2$. Thus, the superfields in the 
$\theta$-twistor formalism preserve their auxiliary fields and give 
an off-shell description of the chiral
 supermultiplets.

Here  we investigate the component structure of the scalar superfield
 $F(\bf\Xi)$ generated by its expansion with respect to the 
both $\bar\nu_{\dot\alpha}$
 and $\bar\eta_{m}$,  and find the structure to be associated with 
an infinite chain of massless chiral supermultiplets $(S, 
S+\frac{1}{2})$ with spin $S=0, 1/2,1,3/2, 2,... \, $. The chain 
includes the well-known massless scalar $\left(0, 
\frac{1}{2}\right)$, vector $\left(\frac{1}{2},1\right)$ and other 
higher spin massless supermultiplets previously  studied in 
\cite{OS,F,FV,BHNW,GS,dWvH,KS,GK} and many other papers. We prove 
that the $\bar\eta_{m}$ expansion of generalized chiral superfields 
in the $\theta$-twistor superspace turns out to be equivalent to the 
power series expansion in $\theta_{\alpha}$.

\section {The Penrose twistor}

To present  the $\theta$-twistor construction \cite{Z} in a clearer 
form we start here from the complexified Minkowsky space with its  
Lorentz group locally isomorphic to $SL(2C)\times SL(2C)$. The 
positive chirality  Weyl spinor $\nu_{\alpha}$  and its complex 
conjugate $\bar\nu_{\dot\alpha}$ with negative  chirality form the 
fundamental representation $\left(\frac{1}{2}, 0\right)$ and its 
c.c.  $\left(0, \frac{1}{2}\right)$ of  the group. Such spinors have  
been  used by Penrose to construct twistors \cite{PR}. The complex 
coordinates  $z_{m}$ of a point in the complexified Minkowsky space 
are represented by the non-Hermitian  $2\times2$ matrix   
 $z_{\alpha\dot\alpha}\equiv z_{m}(\sigma^{m})_{\alpha\dot\alpha}$, 
where $\sigma^{m}=(1,\vec\sigma)$, with $\vec\sigma$ being the $2\times2$ Pauli spin matrices \cite{WB}.
  For our objective it is convenient to introduce the Penrose  
twistors starting from a subspace of holomorphic functions $f(z_{\alpha\dot\alpha}, 
\bar\nu_{\dot\alpha})$ which satisfy the constraint \cite{Z}
\begin{equation}\label{tc} 
\bar\nu_{\dot\alpha}\frac{\partial}{\partial z_{\alpha\dot\alpha}}f(z,\bar\nu)=0. 
\end{equation} 
The general solution of (\ref{tc}) is given by arbitrary  functions 
$f( l_{\alpha}, \bar\nu_{\dot\alpha})$ depending on an effective spinor variable $ l_{\alpha}$
 defined by the Penrose incidence relation \cite{PR} 
\begin{equation}\label{inc}
l_{\alpha} - z_{\alpha\dot\alpha} \bar\nu^{\dot\alpha}=0.
\end{equation} 
The relation (\ref{inc}) is invariant under the shifts of  
$z_{\alpha\dot\alpha}$  by the complex null vector 
$\lambda_{\alpha}\bar\nu_{\dot\alpha} $, where $\lambda_{\alpha}$ is 
an arbitrary spinor. When $\bar\nu_{\dot\alpha}$ is fixed  and 
$\lambda_{\alpha}$ varies,
 a complex totally null plane in the  Minkowski space is swept. 
Penrose called such a null plane
 the $\alpha$-plane. One could think about the $\alpha$-plane as the 
worldvolume swept by a null three-brane \cite{Znul}.
The pair of spinors $\bar\nu_{\dot\alpha}, \,l_{\alpha}$ composes the four
 dimensional complex object called the twistor $\Xi_{A}$ or its  c.c. $\bar\Xi^{ A}$
\begin{equation}\label{9}
\Xi_{A}=(-il_{\alpha},\bar\nu^{\dot\alpha}),\quad
\bar\Xi^{ A}\equiv(\Xi_{A})^*=(\nu^{\alpha},i{\bar l}_{\dot\alpha}).
\end{equation}
Penrose indicated that the complex points $z$ are incident with the 
twistor. It means that if $\nu $ and $l$ are fixed  and 
 Eq. (\ref{inc}) is considered as equation for $z$ then its general solution  
 is given by  $z$ belonging to a two-dimensional complex  $\alpha$-plane in four-dimensional 
complex Minkowski space. The twistor $\Xi_{A}$ and its c.c. 
$\bar\Xi^{A}$  (\ref{9}) yield the quadratic Hermitian form 
\begin{equation}\label{qf} 
\begin{array}{c}
\Xi_{A}
\bar\Xi^{A}=i[ -l_{\alpha}\nu^{\alpha}+ \bar\nu^{\dot\alpha}{\bar l}_{\dot\alpha} ]= 
-i\nu^{\alpha}[z_{\alpha\dot\beta}- \bar z_{\dot\beta\alpha}]\bar\nu^{\dot\beta}
=-i\nu(z- z^{\dag})\bar\nu
\end{array}
\end{equation}
that vanishes for the Hermitian $z$-matrices: $z_{\alpha\dot\beta}
={\bar z}_{\beta\dot\alpha}$. The case corresponds to the real 
Minkowski space and respectively converts the twistor to the null 
twistor. The real space-time point $x_{\alpha\dot\alpha}$ defined by 
the Hermitian matrix $z$ is  associated with the real light vector 
$\nu_{\alpha}\bar\nu_{\dot\alpha}$ going through it.
 The light vector is represented by a point on the Riemann sphere interpreated as a projective line
 $CP^{1}$ belonging to a subspace of the null twistors imbedded into the complex projective 3-space $CP^{3}$. 
 Thus, any point in the real space-time is represented by a Riemann sphere in the projective null 
twistor space. The construction described above admits a 
straightforward supersymmetric generalization resulting in the  
$\theta$-twistor.

\section{ Supersymmetry and the $\theta$-twistor}

Supersymmetry implies the extension of the real Minkowski space 
coordinates $
 x_{\alpha\dot\alpha}$ by the Grassmannian spinor coordinates $\theta_\alpha$ and 
$\bar\theta_{\dot\alpha}$. The corresponding  superspace has  the 
coordinates $( x_{\alpha\dot\alpha},\,\theta_\alpha,\, 
\bar\theta_{\dot\alpha})$ and is invariant  under the 
super-Poincar\'e symmetry \cite{WB}. The  superspace may also be  
extended by the addition of the Penrose spinors $\nu_{\alpha}$ and 
$\bar\nu_{\dot\alpha}$. Using the conventions \cite{UZ} we define 
the supersymmetry transformations in the $D=4\, {\cal N}=1$ 
superspace as follows
\begin{equation}\label{1/1}
\begin{array}{c}
\delta\theta_\alpha=\varepsilon_\alpha,\quad 
\delta x_{\alpha\dot\alpha}=
2i(\varepsilon_{\alpha}\bar\theta_{\dot\alpha}-
\theta_{\alpha}\bar\varepsilon_{\dot\alpha}), \quad \delta\nu_\alpha = 0,
\end{array}
\end{equation}
where $ \nu_{\alpha},\,\bar\nu_{\dot\alpha}$ are not transformed. 
The odd
$D^{\alpha},\,{\bar D}^{\dot\alpha}$ and  even $\partial^{\dot\alpha\alpha} \equiv 
\frac{\partial}{\partial x_{\alpha\dot\alpha}}$ derivatives 
 \begin{equation}\label{3/1}
\begin{array}{c}
D^{\alpha}=\frac{\partial}{\partial\theta_\alpha}-2i\bar\theta_{\dot\alpha}
\partial^{\dot\alpha\alpha}, \quad
{\bar D}^{\dot\alpha}\equiv -(D^{\alpha})^{*}=\frac{\partial}{\partial\bar\theta_{\dot\alpha}}- 
2i\theta_{\alpha}\partial^{\dot\alpha\alpha},
 \quad  
\{ D^{\alpha},\bar D^{\dot\alpha}\}=-4i\partial^{\dot\alpha\alpha}
\end{array}
\end{equation}
in the superspace are also invariant under the transformations (\ref{1/1}). 
A  function $F(x,\theta,\bar\theta)$ in the superspace is known as a superfield. 
 The superfields satisfying the chiral constraint 
\begin{equation}\label{61/1}
\begin{array}{c}
{\bar D}^{\dot\alpha}F(x,\theta,\bar\theta)=0 \longrightarrow 
F=F(y,\theta), \quad 
y_{\alpha\dot\alpha}=x_{\alpha\dot\alpha}-2i\theta_{\alpha}\bar\theta_{\dot\alpha}
\end{array}
\end{equation}
are important for the supersymmetric Yang-Mills theory, supergravity and superstring.
The general solution of  (\ref{61/1}) is given by a chiral superfield
 $F=F(y,\theta)$ depending on the complex coordinates $y_{\alpha\dot\alpha}$  whose  
 imaginary part is the nilpotent monomial 
(-$2\theta_{\alpha}\bar\theta_{\dot\alpha}$). The subspace 
 $(y_{\alpha\dot\alpha},\theta_{\alpha})$  called the {\it chiral} superspace \cite{WB} is closed under 
the supersymmetry transformations 
\begin{equation}\label{10/1}
\delta\theta_\alpha=\varepsilon_\alpha,
\quad 
\delta y_{\alpha\dot\alpha}=
-4i\theta_{\alpha}\bar\varepsilon_{\dot\alpha}
 \end{equation}
and  preserves  the Cartan-Volkov differential one-form $\omega_{\alpha\dot\alpha}$  
\begin{equation}\label{11/1} 
\omega_{\alpha\dot\alpha}=dy_{\alpha\dot\alpha}+
4id\theta_{\alpha}\bar\theta_{\dot\alpha}, \quad \delta \omega_{\alpha\dot\alpha}=0.
\end{equation} 
 After transition to the chiral space the $\theta$-twistor is introduced
by a natural generalization of the above considered twistor construction. It implies the  extension
 of any  chiral superfield $ F=F(y,\theta)$ to a generalized chiral 
superfield $F(y,\theta,\bar\nu)$ depending on the new argument $\bar\nu$. The extension allows to 
 generalize the  constraint (\ref{tc}) to the supersymmetric one
\begin{equation}\label{64/1'} 
\begin{array}{c}
\bar\nu_{\dot\alpha}\frac{\partial}{\partial 
y_{\alpha\dot\alpha}}F(y, \bar\nu, \theta)=0  \quad \longrightarrow 
\quad  F=F( l_{\alpha}, \bar\nu_{\dot\alpha}, \theta_{\alpha})\,,
\end{array} 
\end{equation} 
consistent with the chiral constraint (\ref{61/1}). The constraint (\ref{64/1'}) is satisfied by 
any  chiral superfield $F( l_{\alpha}, \bar\nu_{\dot\alpha}, \theta_{\alpha})$ depending on the 
new spinor $l_{\alpha}$ defined by the generalized ${\it incidence \, relation}$
\begin{equation}\label{sinc}
l_{\alpha} - y_{\alpha\dot\alpha} \bar\nu^{\dot\alpha}=0.
\end{equation} 
The transformations of the spinor $l_{\alpha}$ (\ref{sinc}) under the supersymmetry (\ref{10/1}) 
are nonlinear
\begin{equation}\label{8}
{\delta l}_{\alpha}=-4i\theta_{\alpha}(\bar\nu^{\dot\beta}\bar\varepsilon_{\dot\beta}),\quad
\delta\theta_{\alpha}=\varepsilon_{\alpha}, \quad
\delta\bar\nu_{\dot\alpha}=0
\end{equation}
and reveal the spinor triple $l_{\alpha},\bar\nu_{\dot\alpha},\theta_{\alpha}$ as a new 
representation 
of the supersymmetry.  
The transformations (\ref{8}) are presented as linear by the transition to 
the  new superpartner  $(\theta_{\alpha}\bar\nu_{\dot\beta})$  of $l_{\alpha}$
\begin{equation}\label{8R}
{\delta l}_{\alpha}=4i(\theta_{\alpha}\bar\nu_{\dot\beta})
\bar\varepsilon^{\dot\beta}, \quad
\delta(\theta_{\alpha}\bar\nu_{\dot\beta})=\varepsilon_{\alpha}\bar\nu_{\dot\beta}, \quad
\delta\bar\nu_{\dot\alpha}=0
\end{equation}
 which is the Lorentz vector. 
The equivalent  linear form of the transformations (\ref{8R}) is 
 \begin{equation}\label{8'}
{\delta l}_{\alpha}=-4i(\sigma_{m}\bar\varepsilon)_{\alpha}\bar\eta^{m}, \quad
\delta\bar\eta_{m}=-\frac{1}{2}(\varepsilon\sigma_{m}\bar\nu), \quad
\delta\bar\nu_{\dot\alpha}=0.
\end{equation} 
The Grassmannian vector $\bar\eta_{m}$ in (\ref{8'}) and its c.c $\eta_{m}$ are 
the {\it composite} Ramond vectors  
\begin{equation}\label{vz}
\begin{array}{c} 
\eta_{m}\equiv -\frac{1}{2}(\nu\sigma_{m}\bar\theta),\quad
\bar\eta_{m}=(\eta_{m})^{*}= -\frac{1}{2}(\theta\sigma_{m}\bar\nu),
\\[0.2cm]
\nu_{\beta}\bar\theta_{\dot\alpha}\equiv \eta_{\beta\dot\alpha}=
(\sigma^{m})_{\beta\dot\alpha}\eta_{m},\quad  \eta_{m}\eta_{n}+\eta_{n}\eta_{m}=0
\end{array}
\end{equation} 
 introduced in \cite{VZ} (see details in \cite{Z2}) 
to prove the equivalence between superparticles and spinning particles based on the 
observation that $\bar\eta_{m}$ (\ref{vz}) solves the Dirac constraint 
\begin{equation}\label{id}
\bar\eta_{m}(\bar\nu\tilde\sigma^{m}\lambda)=0,
\end{equation}
where $(\bar\nu\tilde\sigma^{m}\lambda)$ is the tangent vector of the Penrose  $\alpha$-plane.

We see that the supersymmetrization of the complex Minkowski space, 
matching  $y_{\alpha\dot\alpha}$ with $\theta_{\alpha}$, yields the 
supersymmetrization of the Penrose twistor  by the matching  of
$l_{\alpha}$ with $\bar\eta_{m}$. 

As a result, the incidence relation (\ref{inc}) 
gets its Grassmannian counterpart
\begin{equation}\label{svz}
\begin{array}{c}                            
l_{\alpha} - y_{\alpha\dot\alpha} \bar\nu^{\dot\alpha}=0,                   
\\[0.2cm]
\bar\eta_{m} + \frac{1}{2}(\theta\sigma_{m}\bar\nu)=0.                  
\end{array}
\end{equation} 
Thus, the Penrose twistor space $CP^{3}$ is extended up to the superspace 
 $CP^{3|2_{spin}}$ by the addition of the fermionic sector presented by two independent 
complex components of the composite  vector $\bar\eta_{m}$ 
associated with $\theta_{\alpha}$\footnote {We use the notation  
$2_{spin}$ to show the {\it spinor} structure of the  
 fermionic sector of  $CP^{3|2_{spin}}$. The standard notation  $CP^{3|{\cal N}}$ 
of the twistor superspace shows that its fermionic sector is 
represented by the Lorentz {\it scalars} belonging to the 
fundamental representation of the {\it internal} $SU({\cal N})$ 
symmetry of the Yang-Mills theory \cite{Witt2}.}. It is  a 
consequence of the composite structure of $\bar\eta_{m}$  
(\ref{vz}). To see it one can use the spinor basis associated  with 
the supersymmetric triple (\ref{8'}). The basis is formed by a  
Newman-Penrose dyad $\nu^{\alpha}$ and $v^{\alpha}$ \cite {PR} whose 
components satisfy the condition  $\nu^{\alpha}v_{\alpha}=1$. Then 
$\theta_{\alpha}$ may be decomposed in the dyad basis 
\begin{equation}\label{dbe}
\theta_{\alpha}= \eta v_{\alpha}- \chi\nu_{\alpha}, 
\quad \eta\equiv (\nu^{\alpha}\theta_{\alpha}),
\quad  \chi \equiv (v^{\alpha}\theta_{\alpha}),
\end{equation} 
where the complex numbers $\eta $ and $\chi$ are the $\theta$ 
coordinates.   So, the spinor $\theta_{\alpha}$ is  equivalently 
represented by two complex numbers, $\eta$ and $\chi$. The 
substitution of the $\theta$-decomposition (\ref{dbe}) in the
 definition of $\bar\eta_{m}$ 
(\ref{vz}) yields its decomposition
\begin{equation}\label{rvd}
\bar\eta_{m}= -\frac{1}{2}[ \eta (v\sigma_{m}\bar\nu) - \chi(\nu\sigma_{m}\bar\nu)]
\end{equation} 
in  only two basis vectors $(v\sigma_{m}\bar\nu)$ and 
$(\nu\sigma_{m}\bar\nu)$ out of the complete  Newman-Penrose {\it vector}
 basis constructed from the Newman-Penrose dyad  $\nu_{\alpha},\,v_{\alpha}$ 
and their complex conjugate (see details in \cite{GuZ,LiZ}). In this 
vector basis
 originating from the spinor  $\bar\nu^{\dot\alpha}$, belonging to the 
supermultiplet (\ref{8'}),
  the composite Ramond vector  $\bar\eta_{m}$ (\ref{vz}) turns out to be
 represented by the same pair of the complex numbers $\nu$ and $\chi$ as 
 the spinor $\theta_{\alpha}$ (\ref{dbe}). These two complex 
numbers  form the fermionic sector of  $CP^{3|2_{spin}}$. The 
composite structure of $\bar\eta_{m}$ puts severe restrictions on 
its monomials. The monomials of degree greater than two vanish 
because they are
 proportional  monomials formed by the Weyl spinor $\theta_{\alpha}$ 
(see also Eq. (\ref{73/bil})). Such a vanishing never occurs for an arbitrary 
Grassmannian vector $\psi_{m}$, because its  maximal
 monomial $\psi_{0}\psi_{1}\psi_{2}\psi_{3}$ has the degree equal four.

Thus  we obtain the generalization of the twistor (\ref{9}) that  includes the Ramond vector 
 $\bar\eta_{m}$. The corresponding supersymmetric triple has the following 
 form:
\begin{equation}\label{sqrf}
\begin{array}{c} 
{\bf\Xi_{\cal A}}
\equiv(-il_{\alpha},\bar\nu^{\dot\alpha},2\sqrt{2}\bar\eta_{m}),\quad
{\bf\bar\Xi^{\cal A}}\equiv({\bf\Xi}_{\cal A})^*=
(\nu^{\alpha},i{\bar l}_{\dot\alpha},2\sqrt{2}\eta^{m})\,.
\end{array}
\end{equation}
 The triple, called the $\theta$-twistor, yields a new supersymmetric generalization of 
isotropic Penrose twistor. One can check that the quadratic Hermitian  form in the 
$\theta$-twistor superspace 
 \begin{equation}\label{10'}
\begin{array}{c} 
{\bf\Xi_{\cal A}}
{\bf\bar\Xi'^{\cal A}}\equiv i[-l_{\alpha}\nu'^{\alpha}+ \bar\nu^{\dot\alpha}{\bar l'}_{\dot\alpha} 
- 8i\bar\eta_{m}\eta'^{m}],
\\[0.2cm]
{\bf\bar\Xi'^{\cal A}}\equiv({\bf\Xi}'_{\cal A})^*=
(\nu'^{\alpha},i{\bar l}'_{\dot\alpha},2\sqrt{2}\eta'^{m}),
\quad
{\bar l'}_{\dot\alpha}={\bar y}_{\dot\alpha\alpha}\nu'^{\alpha}, \quad 
\eta'_{m}= -\frac{1}{2}(\nu'\sigma_{m}\bar\theta),
\end{array}
\end{equation}
built of coordinates of points ${\bf\Xi_{\cal A}}$ and 
${\bf\bar\Xi'^{\cal A}}$, vanishes (like the form (\ref{qf})) 
because of the supersymmetric incidence relations (\ref{svz}). Next 
we compare the $\theta$-twistor with the supertwistor \cite{Fbr}.

\section{The supertwistor}

The presented derivation of the $\theta$-twistor may be applied for the supertwistor derivation.
A minor change is to start from an equivalent definition of the twistor using the constraint 
\begin{equation}\label{tc1} 
\nu_{\alpha}\frac{\partial}{\partial z_{\alpha\dot\alpha}}f(z,\nu)=0,
\end{equation} 
which differs from (\ref{tc}) by  the substitution of  $\nu_{\alpha}$ for $\bar\nu_{\dot\alpha}$.
The general solution of (\ref{tc1}) is given by the functions 
$f({\bar q}_{\dot\alpha}, \nu_{\alpha})$ depending on the new spinor ${\bar q}_{\dot\alpha}$ 
 defined by the incidence relation 
\begin{equation}\label{inc1}
{\bar q}_{\dot\alpha} - \nu^{\alpha}z_{\alpha\dot\alpha}=0.
\end{equation} 
The comparison of the complex conjugate of Eq. (\ref{inc1}) with (\ref{inc}) shows their  independence
\begin{equation}\label{compr}
\begin{array}{c} 
q_{\alpha} - (\bar\nu\bar z)_{\alpha}=0, 
\\[0.2cm]
l_{\alpha} - (z\bar\nu)_{\alpha}=0 
\end{array}
\end{equation} 
in general and their coincidence for the case of Hermitian  matrices 
$z$: $\bar z = z^{T}$. 

Because the chiral superspace has the non-Hermitian coordinate 
matrix $y_{\alpha\dot\alpha}$,
 the supersymmetric generalizations of the two coincidence relations (\ref{compr}) will not be 
equivalent and in fact they will generate the supertwistor and 
$\theta$-twistor respectively.  In fact, the incidence condition 
(\ref{inc1}) defines another twistor $ Z_{A}$ 
\begin{equation}\label{91}
\begin{array}{c}
Z_{A}=(-iq_{\alpha},\bar\nu^{\dot\alpha}),
\quad
\bar Z^{ A}\equiv(Z_{A})^*=(\nu^{\alpha},i\bar q_{\dot\alpha})
\end{array}
\end{equation}
which include the spinor $q_{\alpha}$ instead of $l_{\alpha}$  (\ref{9}). 

To construct the supersymmetric generalization of the twistor 
(\ref{91}) we go back to the chiral superfields  
$F=F(y,\bar\nu,\theta)$, but replace their argument 
$\bar\nu_{\dot\alpha}$ by
  $\nu_{\alpha}$ and,  respectively,  the constraint (\ref{64/1'}) by the  constraint 
\begin{equation}\label{64/11} 
\begin{array}{c}
\nu_{\alpha}\frac{\partial}{\partial y_{\alpha\dot\alpha}}F(y,\nu, \theta)=0  \quad
\longrightarrow \quad  F=F({\bar q}_{\dot\alpha}, \nu_{\alpha}, \theta_{\alpha})
\end{array} 
\end{equation} 
 which is also supersymmetric and consistent with the chiral constraint (\ref{61/1}). 
The general solution of the constraint (\ref{64/11}) is given by the chiral superfields 
$F( {\bar q}_{\dot\alpha}, \nu_{\alpha}, \theta_{\alpha})$ depending on  the new 
 spinor ${\bar q}_{\dot\alpha}$ 
defined by the incidence relation   (\ref{inc1}) with $y_{\alpha\dot\alpha}$ 
substituted  for  $z_{\alpha\dot\alpha}$
\begin{equation}\label{ssinc}
{\bar q}_{\dot\alpha}- \nu^{\alpha} y_{\alpha\dot\alpha}=0.
\end{equation} 
The transformations of the spinor ${\bar q}_{\dot\alpha}$ (\ref{ssinc}) under the 
supersymmetry (\ref{10/1})  are nonlinear
\begin{equation}\label{8s}
{\delta\bar q}_{\dot\alpha}=-4i(\nu^{\beta}\theta_{\beta})\bar\varepsilon_{\dot\alpha},\quad
\delta\theta_{\alpha}=\varepsilon_{\alpha}, \quad
\delta\nu_{\alpha}=0.
\end{equation}
However, the  transformations (\ref{8s}) may be easily brought to 
the linear form 
\begin{equation}\label{4s}
{\delta \bar q}_{\dot\alpha}=-4i\eta\bar\varepsilon_{\dot\alpha}, \quad
\delta\eta=\nu^{\alpha}\varepsilon_{\alpha}, \quad
\delta\nu_{\alpha}=0, 
\end{equation}
after the contraction of the  second equation in (\ref{8s}) with 
$\nu^{\alpha}$ and transition to the {\it scalar} variable $\eta$ 
(\ref{dbe}), representing only half of the spinor $\theta_{\alpha}$ 
components. The second part of the $\theta$-components described by  
the  complex number  $\chi$ is lost under the transition. 

As a result, the Grassmannian counterpart  
of the incidence relation (\ref{ssinc}) takes the form
\begin{equation}\label{ssinc0}
\eta - (\nu^{\alpha}\theta_{\alpha})=0
\end{equation}
and we obtain the supersymmetrical $\it{ incidence \, relations}$ associated 
with the spinor ${\bar q}_{\dot\alpha}$
\begin{equation}\label{ssinc1}
\begin{array}{c}
{\bar q}_{\dot\alpha}- \nu^{\alpha} y_{\alpha\dot\alpha}=0, 
\\[0.2cm]
\eta - (\nu^{\alpha}\theta_{\alpha})=0,
\end{array} 
\end{equation}
previously obtained in  \cite{Fbr}. These  incidence relations 
generate the 
 supersymmetric triples $Z_{\cal A}$ and $\bar Z^{\cal A}$ unifying 
 $\nu_{\alpha},\bar\nu_{\dot\alpha}$ with  $q_{\alpha}, \bar q_{\dot\alpha}, \eta, \bar\eta$  
\begin{equation}\label{4}
\begin{array}{c}
Z_{\cal A}\equiv(-iq_{\alpha},\bar\nu^{\dot\alpha}, 2\bar\eta),
\quad
\bar Z^{\cal A}\equiv(\nu^{\alpha},i\bar q_{\dot\alpha}, 2\eta).
\end{array}
\end{equation}
The triples $Z_{\cal A}$ and $\bar Z^{\cal A}$ coincide with the  $ 
D=4 \quad {\cal N}=1 $ {\it supertwistor} \cite{Fbr} and its c.c., 
realizing the well-known supersymmetric generalization of the 
projective Penrose twistor.   

The supersymmetric Hermitian quadratic form  \cite{Fbr} in the supertwistor space 
\begin{equation}\label{5}
\begin{array}{c}
 Z_{\cal A}\bar Z'^{\cal A}= i[-q_{\alpha}\nu'^{\alpha}+ \bar\nu^{\dot\alpha}\bar q'_{\dot\alpha} 
-4i\bar\eta\eta'],
\\[0.2cm]
\bar Z'^{\cal A}\equiv  (Z'_{\cal A})^*=(\nu'^{\alpha},i\bar q'_{\dot\alpha}, 2\eta'),
\quad 
\bar q'_{\dot\alpha}=\nu'^{\alpha}y_{\alpha\dot\alpha}, \quad 
\eta'=
\nu'^{\alpha}\theta_{\alpha},
\end{array}
\end{equation}
 where  $Z_{\cal A}$ and  $\bar Z'^{\cal A}$ describes different points of the twistor space, 
vanishes after using the incidence relations (\ref{ssinc1}).
Because of (\ref{ssinc1}) and  (\ref{svz}),  the nonlinear  relations 
\begin{equation}\label{nc}
\begin{array}{c} 
l_{\alpha}=q_{\alpha} - 4i\theta_{\alpha}\bar\eta  , 
\\[0.2cm]
4i\bar\eta\eta'=-4i(\nu'_{\alpha}\bar\nu_{\dot\alpha})\theta^{\alpha}\bar\theta^{\dot\alpha}
=2i(\bar\nu\tilde\sigma_{m}\theta)(\nu'\sigma^{m}\bar\theta)=-8i\bar\eta_{m}\eta'_{m}
\end{array}
\end{equation} 
are fulfilled  and result in the equality of the quadratic forms (\ref{10'}) and (\ref{5})  
\begin{equation}\label{equl}
\begin{array}{c} 
{\bf\Xi_{\cal A}}
{\bf\bar\Xi'^{\cal A}}|_ {(\ref{svz})}= Z_{\cal A}\bar Z'^{\cal A}|_ {(\ref{ssinc1})}
\end{array}
\end{equation}
on the hypersurfaces of the corresponding incidence relations.

Thus, we established  that the extension of  the complex chiral superespace 
$(y_{\alpha\dot\alpha},\theta_{\alpha})$
 by the Penrose  
 spinor $\nu_{\alpha}$, having {\it the same} chirality as $\theta_{\alpha}$,
 generates the supertwistor \cite {Fbr} and the projective superspace $CP^{3|1}$. 
On the contrary,  the extension of  the same  chiral space  by the Penrose  
 spinor $\bar\nu_{\dot\alpha}$ having {\it the\, opposite} chirality to   $\theta_{\alpha}$ 
 yields the $\theta$-twistor \cite{Z} and the projective superspace $CP^{3|2_{spin}}$. 
The principal new property of the $\theta$-twistor is that its 
fermionic  sector forms a {\it vector} representation of the Lorentz 
group in contrast to the supertwistor, whose fermionic sector is 
represented by a Lorentz {\it scalar}. In view of this difference 
the quadratic form (\ref{10'}) is invariant only under the maximal
 subgroup of the superconformal group formed by the supersymmetry (\ref{8'}), the scaling and phase symmetries 
 \begin{equation}\label{52/2}
\begin{array}{c}
l'_{\beta}=e^{\varphi}l_{\beta}, 
\quad
{\bar l}'_{\dot\beta}=e^{\varphi*}{\bar l}_{\dot\beta},
 \quad
\nu'_{\beta}=e^{-\varphi}\nu_{\beta} ,
\quad 
\bar\nu'_{\dot\beta}=e^{-\varphi*}\bar\nu_{\dot\beta},\quad
\\[0.2cm]
\theta'_{\beta}=e^{\varphi}\theta_{\beta}, 
\quad
\bar\theta'_{\dot\beta}=e^{\varphi*}\bar\theta_{\dot\beta},
\quad
\bar\eta'_{m}=e^{2i\varphi_{I}}\bar\eta_{m},\quad
\eta'_{m}=e^{-2i\varphi_{I}}\eta_{m}
\end{array}
\end{equation}
 described  by the complex parameter $\varphi=\varphi_{R}+i\varphi_{I}$, as well as the $\gamma_5$ rotations
 \begin{equation}\label{50}
 \theta'_{\beta}=e^{i\lambda}\theta_{\beta}, \quad 
\bar\theta'_{\dot\beta}=e^{-i\lambda}\bar\theta_{\dot\beta}. 
\end{equation}
The Hermitian form (\ref{10'}) is not invariant under the superconformal boosts  $S^{\alpha}$
 and $\bar S^{\dot\alpha}$  as it follows from the  superconformal boost transformations \cite{Z} 
of the coordinates $y_ {\alpha\dot\alpha},\,\theta_{\alpha}$
\begin{equation}\label{36/1}
\delta y_{\alpha\dot\alpha}=4i\theta_{\alpha}(\xi^{\beta}y_{\beta\dot\alpha}),\quad 
\delta\theta_{\alpha}=- y_{\alpha\dot\beta}\bar\xi^{\dot\beta} + 
4i\theta_{\alpha}(\xi^{\beta}\theta_{\beta})
\end{equation}
 forming the base chiral superspace.
 In fact,  the chiral index $\dot\beta$  of $y_{\alpha\dot\beta}$ in $\delta\theta_{\alpha}$  
(\ref{36/1}) is contracted with the index of the transformation parameter $\bar\xi^{\dot\beta}$.
 Thus, it is not possible  to transform  $y_{\alpha\dot\beta}$ into $l_{\alpha}$, belonging to 
 the $\bf\Xi_{\cal A}$-triple (\ref{sqrf}), by the  contraction of $y_{\alpha\dot\beta}$ with 
$\bar\nu^{\dot\beta}$. On the contrary,  
$y_{\alpha\dot\beta}$ can be contracted  with $\nu^{\alpha}$ to be transformed  into
 ${\bar q}_{\dot\beta}$ 
belonging 
to the $\bar Z^{\cal A}$ triple (\ref{4}). 

We see that  the  difference in the {\it chiralities} of  the $\bar\nu$ and 
 $\theta$ spinors  forming the triple  $ \bf\Xi_{\cal A}$ obstructs the superconformal boost 
realization by the $\theta$-twistor. 

 \section{Massless chiral supermultiplets of higher spin fields}

The superfields $F(\bar Z^{\cal A})$ and $F(\bf\Xi_{\cal A})$ describe  massless supermultiplets
 because they satisfy the Klein-Gordon equations 
\begin{equation}\label{67/1}
\partial_{m}\partial^{m}  F(\bar Z)=0, \quad \partial_{m}\partial^{m} F({\bf\Xi})=0, 
 \end{equation}
where $\partial_{m}\equiv(\sigma_{m})_{\dot\alpha\alpha}\partial^{\dot\alpha\alpha}
\equiv(\sigma_{m})_{\alpha\dot\alpha}\frac{\partial}{\partial x_{\alpha\dot\alpha}},\quad 
\partial^{\dot\alpha\alpha}=-\frac{1}{2}\tilde\sigma_{m}^{\dot\alpha\alpha}\partial^m$. 
 It follows  from the identities
 $\sigma_{m\alpha\dot\alpha}\tilde\sigma^{m\beta\dot\beta}=-2\delta_{\alpha}^{\beta}
\delta_{\dot\alpha}^{\dot\beta}$ and $\bar\nu^{\dot\alpha}\bar\nu_{\dot\alpha}=0$.
The holomorphic superfield $F(\bf\Xi_{\cal A})$ depending on the $\bf\Xi_{\cal A}$-triple is
 expanded in the finite power series in the Grassmannian vector $\bar\eta_{m}$
\begin{equation}\label{73/R}
\begin{array}{c}
 F({\bf\Xi}_{\cal A})\equiv F(-il_{\alpha},\bar\nu^{\dot\alpha}, 2\sqrt{2}\bar\eta_{m})
\\[0.2cm]
=f_{0}(-iy_{\beta\dot\beta}\bar\nu^{\dot\beta},  \bar\nu^{\dot\beta})+ \bar\eta_{m}
f^{m}(-iy_{\beta\dot\beta}\bar\nu^{\dot\beta}, \bar\nu^{\dot\beta}) +\bar\eta_{m}\bar\eta_{n}
f^{nm}_{2}(-iy_{\beta\dot\beta}\bar\nu^{\dot\beta}, \bar\nu^{\dot\beta}).
\end{array}
\end{equation}
The monomial $\bar\eta_{m}\bar\eta_{n}\bar\eta_{l}$ of the third 
degree and higher monomials vanish in view of the composite 
structure of $\bar\eta_{m}$, defined by the incidence relation 
(\ref{svz}) and discussed in the previous section.
 As a result,  the monomials constructed of  $\bar\eta_{m}$ have to be proportional 
to the monomials constructed of the spinor $\theta_{\alpha}$. The 
direct calculation of the $\bar\eta_{m}$-monomials using  the 
definition (\ref{svz}) and the spinor algebra \cite{WB} yields the 
desired relations 
\begin{equation}\label{73/bil}
\begin{array}{c}
\bar\eta_{m}=-\frac{1}{2}(\theta\sigma_{m}\bar\nu),\quad 
\bar\eta_{m}\bar\eta_{n}\equiv\frac{1}{2}[\bar\eta_{m}\bar\eta_{n}-\bar\eta_{n}\bar\eta_{m}]=
\frac{1}{4}(\bar\nu\tilde\sigma_{mn}\bar\nu)\theta^{2},\quad 
\\[0.2cm]
\bar\eta_{m}\bar\eta_{n}\bar\eta_{l}= \theta_{\alpha}\theta_{\beta}\theta_{\gamma}=0,
\end{array}
\end{equation}
where $\theta^{2}\equiv \theta^{\alpha}\theta_{\alpha}$. Because of 
the correspondence (\ref{73/bil}) one can rename the functions in 
(\ref{73/R}) 
\begin{equation}\label{73/rdf}
\begin{array}{c}
\bar\eta_{m}f^{m}=-\frac{1}{2}(\theta\sigma_{m}\bar\nu)f^{m}\equiv-2\theta_{\lambda}f^{\lambda},
\\[0.2cm]
\bar\eta_{m}\bar\eta_{n}f^{nm}_{2}=\frac{1}{4}\theta^{2}(\bar\nu\tilde\sigma_{mn}\bar\nu)f^{nm}_{2}
\equiv\theta^{2}f_{2}
\end{array}
\end{equation}
and present (\ref{73/R}) in the equivalent form of the power series expansion in $\theta_{\alpha}$
\begin{equation}\label{73/1}
 F({\bf\Xi}_{\cal A})\equiv F(-il_{\alpha},\bar\nu^{\dot\alpha},\bar\eta_{m})=
f_{0}(-iy_{\beta\dot\beta}\bar\nu^{\dot\beta},  \bar\nu^{\dot\beta})-
 2\theta_{\lambda}f^{\lambda}(-iy_{\beta\dot\beta}\bar\nu^{\dot\beta}, \bar\nu^{\dot\beta}) +
\theta^{2}f_{2}(-iy_{\beta\dot\beta}\bar\nu^{\dot\beta}, \bar\nu^{\dot\beta}).
\end{equation}
The expansion (\ref{73/1}) has the form of a chiral superfield expansion 
 including the  
{\it auxiliary} field $f_{2}$ that {\it identically\, vanishes} 
in the supertwistor approach \cite{Fbr}. 
It implies the existence of an equivalent  representation  $\tilde F(\Xi_{A})$
 of the superfield  $ F({\bf\Xi}_{\cal A})$
\begin{equation}\label{73/dual}
\begin{array}{c}
 F({\bf\Xi}_{\cal A})=\tilde F(\Xi_{A}),
\end{array}
\end{equation}
which depends on the supersymmetric entirely spinor triple $\Xi_{A}$ 
\begin{equation}\label{ksi}
\Xi_{\cal A}\equiv(-il_{\alpha},\bar\nu^{\dot\alpha},\theta^{\alpha}),\quad
\bar\Xi^{\cal A}\equiv(\Xi_{\cal A})^*=(\nu^{\alpha},i{\bar l}_{\dot\alpha},
\bar\theta^{\dot\alpha})
\end{equation}
associated with the ${\bf\Xi}_{\cal A}$-triple. 
This correspondence between the  ${\bf\Xi}_{\cal A}$ and $\Xi_{A}$ triples
 is the reason to 
use the same name  $\theta$-twistor for the spinor triple 
 $\Xi_{A}$ (\ref{ksi}). 
Consequently, in the following the symbol tilde over the superfields 
depending on $\Xi_{A}$ will be droped.

 We proved that the $\theta$-twistor introduces a natural extension 
 of the base chiral superspace  $(y_{\alpha\dot\alpha},\theta_{\alpha})$ 
and its use preserves all
 components in the chiral superfields depending on $\bar\nu^{\dot\alpha}$. 
 The superfield $\tilde F(\Xi_{A}$) (\ref{73/dual}) depends on the commuting 
spinor $\bar\nu^{\dot\alpha}$ and the coefficents of power expansion 
of $\tilde F(\Xi_{A})$ 
in the spinor degrees are chiral superfields 
$\Phi^{\dot\alpha_{1}...\dot\alpha_{2S}}(y,\theta)$ 
with arbitrary number of dotted spinor indices. To find these 
fields one can 
 use Penrose's idea of the contour integration applied to  
$\tilde F(\Xi_{A}$) that permits to integrate out the $ \bar\nu$ dependence.
The idea was previously extended to the
 supertwistor in \cite{Fbr} and the extension may be applied for the superfields
  depending on the  $\theta$-twistor.
The corresponding contour integral defining the 
superfield $\Phi^{\dot\alpha_{1}...\dot\alpha_{2S}}(y,\theta)$ is given by 
the expression 
\begin{equation}\label{74/1}
\Phi^{\dot\alpha_{1}...\dot\alpha_{2S}}(y,\theta)=
\oint(d\bar\nu^{\dot\gamma}\bar\nu_{\dot\gamma})\bar\nu^{\dot\alpha_{1}}...
\bar\nu^{\dot\alpha_{2S}}
 F(\bar\nu^{\dot\beta},-i\bar\nu^{\dot\gamma}y_{\beta\dot\gamma}, \theta_{\beta}),
\end{equation}
 where we omit the tilde over $F(\Xi)$ (\ref{73/dual}) and assume its homogeneity 
degree equal to $-2(S+1)$. The $\bar\nu$-contour encloses the 
singularities of $F$ in (\ref{74/1}) for each
 fixed point $(y,\theta)$. 
The substitution of  (\ref{73/1}) into (\ref{74/1}) results in the expansion 
\begin{equation}\label{75/1}
\begin{array}{c}
\Phi^{\dot\alpha_{1}...\dot\alpha_{2S}}(y,\theta)=f_{0}^{\dot\alpha_{1}...
\dot\alpha_{2S}}(y)-2\theta_{\lambda}f^{\lambda\dot\alpha_{1}...\dot\alpha_{2S}}(y) +
\theta^{2}f_{2}^{\dot\alpha_{1}...\dot\alpha_{2S}}(y)
\end{array}
\end{equation}
where the component functions are defined  by the integrals
\begin{equation}\label{76/1}
\begin{array}{c}
f_{0}^{\dot\alpha_{1}...\dot\alpha_{2S}}(y)=
\oint(d\bar\nu^{\dot\gamma}\bar\nu_{\dot\gamma})\bar\nu^{\dot\alpha_{1}}...
\bar\nu^{\dot\alpha_{2S}}f_{0}(-iy_{\beta\dot\beta}\bar\nu^{\dot\beta},  \bar\nu^{\dot\beta}),
\\[0.2cm]
f^{\lambda\dot\alpha_{1}...\dot\alpha_{2S}}(y)= 
\oint(d\bar\nu^{\dot\gamma}\bar\nu_{\dot\gamma})\bar\nu^{\dot\alpha_{1}}...
\bar\nu^{\dot\alpha_{2S}}f^{\lambda}(-iy_{\beta\dot\beta}\bar\nu^{\dot\beta}, \bar\nu^{\dot\beta}),
\\[0.2cm]
f_{2}^{\dot\alpha_{1}...\dot\alpha_{2S}}(y)=
\oint(d\bar\nu^{\dot\gamma}\bar\nu_{\dot\gamma})\bar\nu^{\dot\alpha_{1}}...
\bar\nu^{\dot\alpha_{2S}}f_{2}(-iy_{\beta\dot\beta}\bar\nu^{\dot\beta},  \bar\nu^{\dot\beta})
\end{array}
\end{equation}
with $f^{\lambda\dot\alpha_{1}...\dot\alpha_{2S}}(y)$ and  $f_{0,2}^{\dot\alpha_{1}...
\dot\alpha_{2S}}(y)$ satisfying the chiral Dirac equations
\begin{equation}\label{77/1}
\partial_{\alpha\dot\alpha_{k}}
f^{\lambda\dot\alpha_{1}..\dot\alpha_{k}..\dot\alpha_{2S}}(x)= \partial_{\alpha\dot\alpha_{k}}
f_{0,2}^{\dot\alpha_{1}..\dot\alpha_{k}..\dot\alpha_{2S}}(x)=0, 
\quad (k=1,2,...,2s).
\end{equation}
The further expansion of $\Phi^{\dot\alpha_{1}...\dot\alpha_{2S}}(y,\theta)$ 
(\ref{75/1})  at the real point $x_{m}$ is  given by 
\begin{equation}\label{80/1}
\begin{array}{c}
\Phi^{\dot\alpha_{1}...\dot\alpha_{2S}}(y,\theta)=f_{0}^{\dot\alpha_{1}...\dot\alpha_{2S}}(x)
-2\theta_{\lambda}f^{\lambda\dot\alpha_{1}...\dot\alpha_{2S}}(x) -
2i\theta_{\gamma}\bar\theta_{\dot\gamma}\partial^{\dot\gamma\gamma} f_{0}^{\dot\alpha_{1}...
\dot\alpha_{2S}}(x)
\\[0.2cm]
-2i\theta^{2}\bar\theta_{\dot\gamma}\partial^{\dot\gamma\lambda}f_{\lambda}^{\dot\alpha_{1}...
\dot\alpha_{2S}}(x)+\theta^{2}f_{2}^{\dot\alpha_{1}...\dot\alpha_{2S}}(x),
\end{array}
\end{equation}
where the term $\frac{1}{2}\theta^{2}{\bar\theta}^{2}
\partial^{\dot\gamma\gamma}\partial_{\gamma\dot\gamma}f_{0}^{\dot\alpha_{1}...\dot\alpha_{2S}}(x)$ 
was droped because of the zero mass constraint 
(\ref{67/1})
\begin{equation}\label{81/1}
\begin{array}{c}
\Box\Phi^{\dot\alpha_{1}...\dot\alpha_{2S}}(y,\theta)=0 \longrightarrow 
\Box f_{0}^{\dot\alpha_{1}...\dot\alpha_{2S}}(x)=0.
\end{array}
\end{equation}
For sewing these results with the well-known case of the scalar 
supermultiplet, corresponding to $S=0$, we rename the component 
$f$-fields by the generally accepted notations \cite{WB} 
\begin{equation}\label{82/1}
\begin{array}{c}
f_{0}^{\dot\alpha_{1}...\dot\alpha_{2S}}={\sqrt{2}}A^{\dot\alpha_{1}...\dot\alpha_{2S}}
\equiv{\sqrt{2}}A^{...},
\quad
f_{2}^{\dot\alpha_{1}...\dot\alpha_{2S}}=\sqrt{2}F^{\dot\alpha_{1}...\dot\alpha_{2S}}
\equiv\sqrt{2}F^{...},
\\[0.2cm]
f_{\lambda}^{\dot\alpha_{1}...\dot\alpha_{2S}}=\psi_{\lambda}^{\dot\alpha_{1}...\dot\alpha_{2S}} 
\equiv\psi_{\lambda}^{...},
\end{array}
\end{equation}
where $(...)\equiv (\dot\alpha_{1}...\dot\alpha_{2S})$. 
Then we find the superfield $\frac{1}{\sqrt{2}}\Phi^{...}(y,\theta)$ to describe the
 massless chiral 
multiplet \cite{WB} for the case $S=0$. For  $S\neq0$, the 
superfield (\ref{80/1}) represents the chiral
 supermultiplets of massless higher spin fields 
 with the particle spin content $$\left(\frac{1}{2},1\right),\, \left(1, \frac{3}{2}\right),\,
\left(\frac{3}{2},2\right),\, ......., \left(S, 
S+\frac{1}{2}\right)$$  accompanied by the  corresponding  auxiliary 
fields for
 any integer or half-integer 
spin  $ S=\frac{1}{2},1, \frac{3}{2},2, ....$. In  various approaches the
 supermultiplets  were discussed in the papers
\cite{OS,F,FV,BHNW,GS,dWvH,KS,GK}  and many  others. The 
supersymmetry transformations for the higher spin multiplet 
(\ref{80/1}) presented in the notations (\ref{82/1}) take  the form 
\begin{equation}\label{84/1}
\begin{array}{c}
\delta A^{...}=\sqrt{2}\varepsilon^{\lambda}\psi_{\lambda}^{...}, \quad
\delta F^{...}=i\sqrt{2}(\bar\varepsilon{\tilde\sigma}_{m}
\partial^{m}\psi^{...})
\\[0.2cm]
\delta\psi_{\lambda}^{...}=i\sqrt{2}(\sigma_{m}\bar\varepsilon)_{\lambda}\partial^{m}A^{...} + 
\sqrt{2}\varepsilon_{\lambda}F^{...}
\end{array}
\end{equation}
 and coincide  with the transformation rules for the $S=0$ chiral 
multiplet of weight $n=\frac{1}{2}$  \cite{WB}  if we put   
$A^{...}=A,\, F^{...}=F$ and $\psi_{\lambda}^{...}= \psi_{\lambda}$ 
in (\ref{84/1}).

It was above noted that the $\theta$-twistor superspace is invariant under 
the axial rotations (\ref{50}). 
These phase transformations generate  the $R$-symmetry transformations 
for  $\tilde F(\Xi)$ (\ref{73/1})
\begin{equation}\label{85/1}
\begin{array}{c}
 \tilde F'(-il_{\alpha}, \bar\nu^{\dot\alpha},  e^{i\varphi}\theta^{\alpha})=
e^{2in\varphi}\tilde F(-il_{\alpha},
 \bar\nu^{\dot\alpha}, \theta^{\alpha}),
\end{array}
\end{equation}
where $n$ is the correspondent $R$ number. Then  taking into account the representation (\ref{74/1})
 we get 
  the $R$-symmetry transformation of the generalized chiral superfield 
$\Phi^{\dot\alpha_{1}...\dot\alpha_{2S}}(y,\theta)$
\begin{equation}\label{86/1}
\Phi'^{\dot\alpha_{1}...\dot\alpha_{2S}}(y,\theta)=
e^{2in\varphi}\Phi^{\dot\alpha_{1}...\dot\alpha_{2S}}(y,e^{-i\varphi}\theta).
\end{equation}

Thus we conclude, that the transition  to  the $\theta$-twistor from 
the supertwistor permits to solve the off-shell description problem 
for the massless higher spin chiral supermultiplets. The off-shell 
spinor superfield $\Phi^{\dot\alpha}(y,\theta)$ describes the  
${\cal N}=1 \,\, D=4$  gluon and gluino with fixed helicities. The 
complex conjugate superfield describes the  gluon and gluino with 
opposite helicities. Thus, the states with opposite helicities are 
associated 
 with the holomorphic ${\bf\Xi}_{\cal A}$ and the antiholomorphic parts 
${\bf\bar\Xi^{\cal A}}$ of the complete $\theta$-twistor space. The construction
permits a direct generalization to the case of ${\cal N}=4 \,\, D=4$  super Yang-Mills.

 \section{Conclusion} 

The new concept of supersymmetric $\theta$-twistor, alternative to 
the well-known supertwistor, and its physical applications have been 
discussed. The fermionic sector of the $\theta$-twistor is 
represented by the composite
 Grassmannian Lorentz vector  $(\bar\nu\gamma_{m}\theta)$, 
providing the extension of the Penrose projective twistor space to 
the new projective superspace different from the supertwistor space. 
The new property of the proposed supersymmetrization of the Penrose 
twistor consists in the presence of the Lorentz {\it vector} in the 
fermionic sector of the $\theta$-twistor. The supertwistor fermionic 
sector is represented by a Lorentz {\it scalar}. The ${\cal N}=1\,\, 
\theta$-twistor considered here is naturally generalized to include 
the
 internal symmetry, similar to the extension for the supertwistor case.
This is achieved by the substitution of $\theta_{\alpha}^{i}$ for  $\theta_{\alpha}$ in
the $\theta$-twistor components, 
where the index  $i$ belongs to the fundamental representation of the group $SU(N)$. 
 The substitution yields a new composite Grassmannian vector  
$(\bar\nu\gamma_{m}\theta)^{i}$ carrying  both the space-time and internal 
symmetry indices. 
The corresponding  extension of the projective $CP^{3}$  space yields a 
new extended projective space, 
 whose fermionic sector {\it  mixes} the space-time and internal symmetries. 
This property deserves further study along the line developed in \cite{GN}.
The $\theta$-twistor discussed here is not covariant under the 
superconformal boosts contrary to the supertwistor, but it is covariant under 
the maximal subgroup of the superconformal group.
 The superconformal boosts appear to be a broken symmetry 
of the $\theta$-twistor space. This breaking correlates with the 
Gross-Wess effect of the {\it conformal} symmetry breaking in
 the {\it scale} 
invariant amplitudes of the scalar-{\it spinor} and scalar-{\it 
vector} particles scattering \cite{GW}. An attractive property of 
the breaking is that it is accompanied by restoration of the 
auxiliary fields of the chiral supermultiplets, absent in the 
supertwistor description. Having observed this property, we applied 
the $\theta$-twistor for
 the construction 
of the physical supermultiplets describing the ${\cal N}=1 \,\, D=4 \, $ 
physical fields.
As a result, we find infinite chain of massless higher spin ${\cal N}=1 \,\, D=4$ 
 chiral 
supermultiplets $(S, S+1/2)$ {\it containing} their auxiliary $F$ 
fields. The above mentioned extension of the $\theta$-twistor 
superspace to describe the  $SU(N)$ group degrees of freedom shows a 
way of an off-shell description of the  ${\cal N}=4 \,\, D=4$  
super Yang-Mills theory. This  extension has to be accompanied by 
the auxiliary $F$ field restoration, resulting in the superconformal 
symmetry breaking. The breaking is present in the ${\cal N}=1 \,\, 
D=10$  super Yang-Mills theory, but the recent proposal of  
Berkovits \cite{Ber_10} to built its $\theta$-twistor description 
stimulates the  $\theta$-twistor description of ${\cal N}=4 \,\, D=4 
\, $   super
 Yang-Mills.
Another application \cite{Z2} of the  $\theta$-twistor superspace is connected with 
 a possible way to solve the problem of Lorentz symmetry breaking in supersymmetric 
non(anti)commutative geometry, alternative to using the twisted Hopf 
algebra construction recently proposed in \cite{MA1,MA2}. In 
general, it might be that the exact superconformal symmetry is not 
realized in nature. At least we know that the (super)conformal 
gravity is plagued with ghosts when its perturbation quantization is 
considered \cite{KTN,Kaku}. Then the $\theta$-twistor superspace, 
minimally breaking superconformal
 symmetry, but preserving the super Poincar\'e and other important symmetries,  
might be used as a natural base space-time for new supesymmetric model building.

\section{Acknowledgements}

We thank I. Bengtsson, P. Di Vecchia, F. Hassan, P. Howe, S. 
Theisen, K. Tod, P. Townsend, D. Uvarov and J.W. van Holten for 
useful discussions. This work was partially supported by the Grants 
of the Academy of Finland, under
 the  Projects No. 121720 and 127626, and Nordita.

\end{document}